\documentclass{nature}

\usepackage[colorlinks,linkcolor={blue},citecolor={blue},urlcolor={blue}]{hyperref}
\usepackage{threeparttable}
\usepackage{graphicx}
\usepackage{makecell}
\usepackage{amsmath}
\usepackage{amssymb}    
\usepackage{url}        
\usepackage[font={small,stretch=1.25}]{caption}
\usepackage{orcidlink}

\spacing{1.4}

\title{Cosmic CO and [CII] backgrounds and the fueling of star formation over 12 Gyr}

\author{
Yi-Kuan Chiang\orcidlink{0000-0001-6320-261X}$^{1*}$
}

\newcommand{\araa}{Annu. Rev. Astron. Astrophys.}   
\newcommand{\aj}{Astron. J.}   
\newcommand{\apj}{Astrophys. J.}   
\newcommand{\apjs}{Astrophys. J. Suppl. Ser.}   
\newcommand{\aap}{Astron. Astrophys.}   
\newcommand{\aapr}{Astron. Astrophys. Rev.}   
\newcommand{\mnras}{Mon. Not. R. Astron. Soc.}   
\newcommand{\physrep}{Phys. Rep.}   
\newcommand{\prd}{Phys. Rev. D}   
\newcommand{\pasp}{Publ. Astron. Soc. Pac.}   

\begin{document}

\maketitle

\begin{affiliations}
 \item Academia Sinica Institute of Astronomy and Astrophysics (ASIAA), Taipei 106319, Taiwan.\\
 {$^{*}$E-mail: ykchiang@asiaa.sinica.edu.tw}
\end{affiliations}

\begin{abstract}
\boldmath
Molecular gas, modest in mass yet pivotal within the cosmic inventory, regulates baryon cycling as the immediate fuel for star formation. Across most of cosmic history, its reservoir has remained elusive, with only the tip of the iceberg revealed by luminous carbon monoxide (CO) emitting galaxies. Here we report the first detections of the mean cosmic CO background across its rotational ladder at 7$\sigma$, together with ionized carbon ([CII]) at 3$\sigma$, over $0<z<4.2$. This uses tomographic clustering of diffuse broadband intensities with reference galaxies, directly probing aggregate emission in the cosmic web. From CO(1--0) we infer the total molecular gas density, $\Omega_{\rm H_2}$, finding it about twice that resolved in galaxy surveys. The global depletion time is $\sim$1~Gyr, shorter than the Hubble time, requiring sustained inflow. CO excitation links to star-formation surface density and, with depletion time, yields a super-linear Kennicutt--Schmidt law that appears universal. Together these results establish a global picture of galaxy growth fueled by a larger, short-lived molecular reservoir. The CO and [CII] detections mark a turning point for line-intensity mapping, replacing forecasts with empirical line strengths and defining sensitivity requirements for upcoming 3D experiments poised to open new windows on galaxy formation and cosmology.
\end{abstract}

\clearpage

\renewcommand{\figurename}{\bf Fig.}
\renewcommand\thefigure{\arabic{figure}}

Understanding the cosmological growth of galaxies requires probing the gas reservoirs that regulate the baryon cycle within the cosmic ecosystem\cite{2020ApJ...902..111W}. A defining feature of cosmic history is the rise of the star-formation rate (SFR) density to a peak at $z\sim2$, followed by a ten-fold decline to the present\cite{2014ARA&A..52..415M,2025ApJ...992...65C}. The driver of this evolution remains uncertain: is it set by the global gas supply, or by changes in how efficiently galaxies consume it? Line emission from CO and [CII] probes key baryon phases in the interstellar medium (ISM) and circumgalactic medium (CGM), with CO tracing the bulk molecular hydrogen (H$_2$) that fuels star formation and [CII] tracing the cooling medium left in its wake. Beyond the local universe, however, galaxy surveys face challenges in sensitivity and efficiency; existing CO and [CII] programs cover only small areas or pre-selected targets, yielding a few hundred emitters in total and far fewer at any given redshift\cite{2019ApJ...872....7R,2019ApJ...882..138D,2020AJ....159..190L,2020ApJ...902..110D,2023ApJ...945..111B,2018MNRAS.481.1976Z,2020A&A...643A...2B}. Such sparse sampling, coupled with surface-brightness limits, leaves the luminosity functions and total light budgets highly uncertain. These challenges, along with opportunities to probe large-scale structure (LSS) and cosmology, motivate line-intensity mapping (LIM), a new approach that bypasses the need to resolve individual galaxies and instead captures the integrated emission of all sources in the cosmic web.

Our CO and [CII] intensity measurements build on a companion study\cite{2025ApJ...992...65C} of the cosmic infrared background (CIB), where young starlight is absorbed by dust and thermally re-emitted, a process that dominates the energy output of galaxy formation. That analysis used diffuse emission in 11 broadband intensity maps from {\it Planck}, {\it Herschel}, and {\it IRAS}\cite{2005ApJS..157..302M,2012MNRAS.424.1614O,2016A&A...594A...8P,2017ApJS..233...26S}, spanning a 50-fold frequency range from $\rm 100\,GHz$ to $\rm 5\,THz$. Redshift information was obtained through spatial cross-correlations with reference galaxies and quasars (\hyperref[sec_tomo]{Methods}) over an effective sky of $1\pi$ sr, enabling tomographic reconstruction of the CIB spectrum over $0<z<4.2$ robust against Galactic foregrounds. In the large-scale ``clustering’’ regime, gravity links these bright references to the underlying galaxy population rather than to their own emission or the shot noise from discrete bright sources. This enables direct estimates of the mean backgrounds, free from galaxy incompleteness and astrophysical assumptions. Using this clustering-based CIB tomography, the companion paper\cite{2025ApJ...992...65C} traced the evolution of dust temperature and the total dust mass budget $\Omega_{\rm dust}$, recovered the cosmic SFR history with substantially improved precision over previous benchmark\cite{2014ARA&A..52..415M}, and showed that star formation has been predominantly dust-obscured for over 12\,Gyr.

\begin{figure}
\includegraphics[width=1\textwidth]{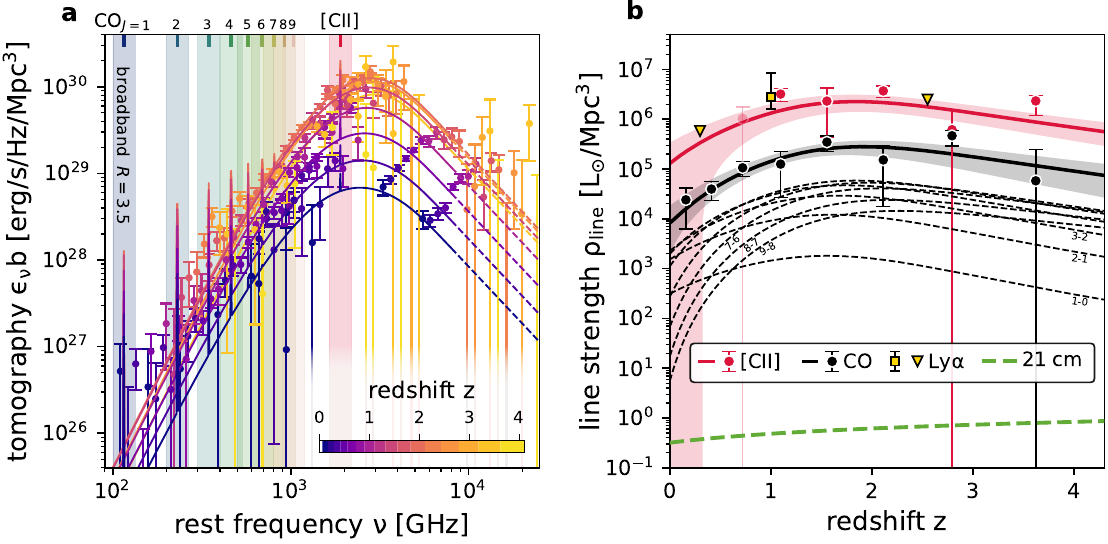}
\centering
\caption{\textbf{Tomographic CIB spectrum and cosmic CO and [CII] histories.}
\textbf{a}, Evolving CIB spectrum shown as bias-weighted emissivity $\epsilon_{\nu}b$ versus rest-frame frequency, color-coded by redshift. Data points are from 11-band intensity-galaxy cross-correlations in the two-halo clustering regime\cite{2025ApJ...992...65C}. Vertical bands mark nine CO and the [CII] 158~$\mu$m lines, with widths set by the mean spectral resolution $R\sim3.5$; data points within these bands receive line excess. Colored curves show the best-fit continuum-plus-line model, with unresolved CO and [CII] drawn as tophats at 1\% of their frequencies. 
\textbf{b}, We detect the cosmic CO and [CII] backgrounds by exploiting their characteristic frequency-redshift line-excess patterns in panel \textbf{a}. The comoving luminosity densities, with CO summed over nine lines (black and red curves with shaded intervals), peak broadly near cosmic noon. Dashed lines give posterior medians for individual CO transitions, revealing evolving excitation and the dominance of mid-$J$ transitions. Black and red points are per-redshift refits, supporting our redshift parameterization; the single transparent [CII] point is negative due to noise. [CII] is unconstrained at $z<0.4$ owing to a gap in spectral coverage. For broader context, yellow data points and upper limits (95\%) indicate Ly$\alpha$ intensity-mapping constraints\cite{2019ApJ...877..150C,2018MNRAS.481.1320C}, and the green dashed line shows predicted hydrogen 21~cm amplitudes\cite{2024ApJ...963...23A}. [CII] likely rivals Ly$\alpha$ as the brightest line background, while CO remains much brighter than 21~cm. All error bars and shaded bands indicate 1$\sigma$ (68\%) uncertainties.}
\label{fig:spectrum_and_lines}
\end{figure}

Here we unlock the full potential of intensity tomography to detect CO and [CII] emission and probe the cosmic budgets of molecular and cooling gas. Fig.~\ref{fig:spectrum_and_lines}a shows the rest-frame bias-weighted mean CIB emissivity spectrum, $\epsilon_{\nu} b(\nu,z)$, from the companion work\cite{2025ApJ...992...65C}, where the emissivity $\epsilon_{\nu}$ is the radiative power per unit frequency and comoving volume in $\rm erg\ s^{-1}\ Hz^{-1}\ Mpc^{-3}$, and $b$ is the effective CIB bias relative to the underlying matter density fluctuations. The broad thermal dust emission traces the obscured SFR history, while the discrete frequencies of the CO rotational ladder and the [CII] 158~$\mu$m line mark where these line contributions enter. Redshift tomography enhances the effective spectral resolution by mapping each observed band $\nu_{\rm obs}$ to $\nu=\nu_{\rm obs}(1+z)$ across 16 redshift bins, yielding super-bandpass sampling for line diagnostics.

We analyze the total emissivity $\epsilon_\nu$ as the sum of continuum and line components (\hyperref[sec_full_model]{Methods}). The continuum follows the ensemble dust spectrum from the companion study\cite{2025ApJ...992...65C}, with parameters describing dust density, temperature, opacity index, and their redshift evolution. The line term includes [CII] 158~$\mu$m at 1900.54~GHz, treated as a Dirac delta function, and nine CO transitions from $J{=}1\text{--}0$ (115.27~GHz) through $J{=}9\text{--}8$. Other lines are fainter and expected to be negligible at our current emissivity precision. The CO ladder is fit jointly with a spectral line energy distribution (SLED) controlled by an excitation parameter linked to an effective galaxy-scale star-formation surface density $\Sigma_{\rm SFR}$, described by equation~(\ref{eq:NK14}) in \hyperref[sec_sled]{Methods} with coefficients given in Extended Data Table~\ref{tab:SLED_coefficients}. We perform Bayesian inference on the tomographic $\epsilon_{\nu} b(\nu,z)$ spectrum in Fig.~\ref{fig:spectrum_and_lines}a, supplemented by {\it FIRAS} and {\it Planck} monopole measurements\cite{2019ApJ...877...40O} to break the bias-intensity degeneracy. This yields redshift-resolved luminosity densities for CO and [CII], $\rho_{\rm CO,\,J(J{-}1)}(z)$ and $\rho_{\rm CII}(z)$, together with the mean $\Sigma_{\rm SFR}(z)$ that governs CO excitation. A summary of all model parameters and posteriors is provided in Extended Data Table~\ref{tab:parameters}.

We detect a statistically significant $6.6\sigma$ cosmic CO signal, quantified by $\rho_{\rm CO}$ summed over nine lines after marginalizing over continuum and accounting for redshift covariance. [CII] is also detected at $3.0\sigma$. Fig.~\ref{fig:spectrum_and_lines}b shows the posterior evolution of $\rho_{\rm CO}$ and $\rho_{\rm CII}$ (thick lines and bands), both peaking at $z\sim2$. For visualization, we refit the data for per-redshift estimates (circular markers), allowing CO(1--0) and [CII] amplitudes to vary across 8 and 6 bins, respectively. Posterior medians for individual CO transitions are also shown, providing excitation diagnostics and further validation of the CO signals. These detections are robust to foregrounds, coeval continuum, and line confusion, representing the first measurements of the [CII] background and of CO beyond the shot-noise regime. They recover the total cosmic mean intensity, encompassing the full galaxy population down to the faintest objects, with potential minor contributions from molecular outflows and extended CGM halos\cite{2019ApJ...887..107F}. Compared to the CIB continuum\cite{2025ApJ...992...65C}, [CII] accounts for about 0.3 percent of the total infrared luminosity density, and the nine CO lines contribute about 0.03 percent. Their redshift evolution tracks the total cosmic SFR more closely than the CIB, implying sensitivity to the full star-forming gas rather than only its dust-obscured component (\hyperref[sec_lines_to_total]{Methods}).

To place CO and [CII] in a broader context of line backgrounds, Fig.~\ref{fig:spectrum_and_lines}b also shows Ly$\alpha$ intensity constraints: a $2\sigma$ detection at $z=1$ and a limit at $z=0.3$\cite{2019ApJ...877..150C}, and another limit at $z=2.55$\cite{2018MNRAS.481.1320C}. Given current uncertainties, the competition for the brightest line is a tie between Ly$\alpha$ and [CII], with [CII] likely pulling ahead at low redshift as metallicity increases and Ly$\alpha$ escape decreases. The predicted hydrogen hyperfine 21\,cm background is significantly fainter\cite{2024ApJ...963...23A}. Notably, [CII], though a ``metal'' line, may outshine all hydrogen lines, making it a premier LSS tracer.

\begin{figure*} \centering
    \includegraphics[width=1\textwidth]{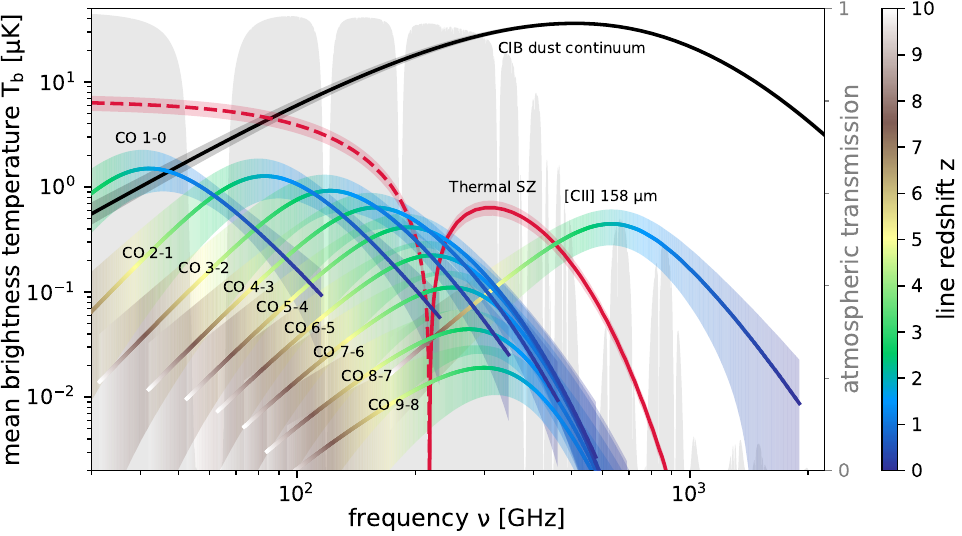}
    \caption{\textbf{Landscape of millimeter line monopoles in brightness temperature.} The mean sky monopoles $T_{\rm b}$ in $\mu$K versus observed frequency for the CO and [CII] backgrounds, derived from our posterior constraints, are shown as curves color-coded by redshift, with shading indicating the 68\% uncertainties. The black curve and band show the updated CIB dust-continuum monopole in this work fitted jointly with the lines, with their sum closely matching the previous continuum-only fit\cite{2025ApJ...992...65C}. The spectral distortion of the thermal Sunyaev-Zeldovich background is shown in red, with the dashed segment indicating the decrement ($y \approx 1.22 \times 10^{-6}$)\cite{2020ApJ...902...56C}. All component amplitudes are now empirically constrained, with the lines tracing LSS in 3D and the continuum tracing it in 2D, informing both LIM analyses and foreground modeling or mitigation for CMB experiments. Filled gray regions mark atmospheric transmission windows at a relatively dry site\cite{2001ITAP...49.1683P}, highlighting more accessible low-frequency bands from the ground. This figure establishes key benchmarks for the far-infrared to millimeter line backgrounds, replacing previously divergent model forecasts with direct measurements and defining amplitude baselines and sensitivity targets for future LIM experiments.}
\label{fig:T_b}
\end{figure*}

Fig.~\ref{fig:T_b} summarizes the sky monopole, i.e., the mean brightness temperatures $T_{\rm b}$ of the CO and [CII] line backgrounds, converted from the luminosity densities in Fig.~\ref{fig:spectrum_and_lines}b. An intensity version is shown in Extended Data Fig.~\ref{fig:I_nu}. Each line appears at its rest frequency at $z=0$ and redshifts to lower frequencies, extrapolated to $z=10$, with shaded bands indicating the $1\sigma$ uncertainties. For comparison, we overlay the jointly constrained dust continuum from this work and the empirical estimate of the thermal Sunyaev-Zeldovich background\cite{2020ApJ...902...56C}, whose decrement exceeds the dust emission below $\sim80$~GHz. Fainter lines, as well as the radio free-free and synchrotron backgrounds rising at the lowest frequencies, are neglected. Although [CII] is far brighter in absolute terms, CO lines become increasingly prominent at lower frequencies, showing substantially higher equivalent widths (\hyperref[sec_equivalent_width]{Methods}) relative to the CIB and other secondary effects of the cosmic microwave background (CMB).

\begin{figure}
\includegraphics[width=0.775\textwidth]{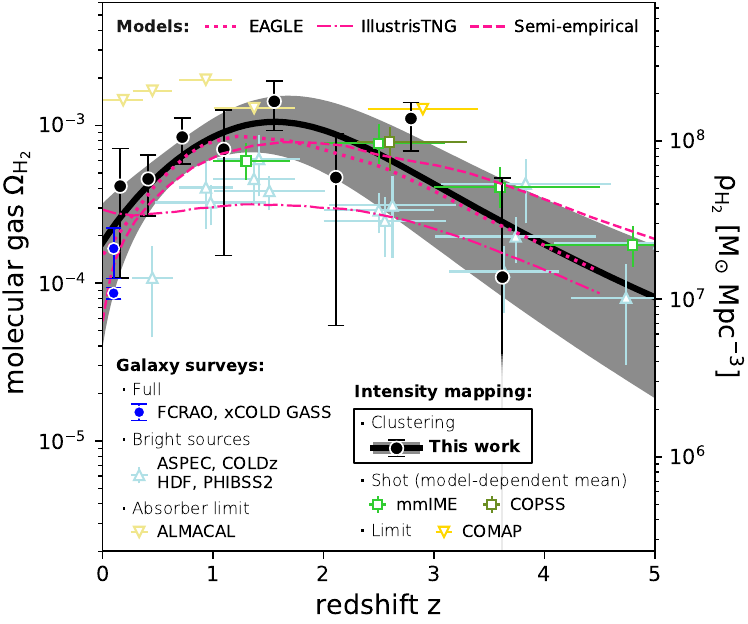}
\centering
\caption{\textbf{History of molecular gas inventory traced by CO.} Expressed as the molecular density parameter $\Omega_{\rm H_2} \equiv \rho_{\rm H_2} / \rho_{\rm crit}$ evolution, our intensity-based measurement is from the CO(1--0) background, jointly constrained by all nine transitions over $z = 0$--4.2. Results are shown as the black line (posterior median) with shaded 68\% credible intervals and black data points for per-redshift re-fits. Modulo CO-to-H$_2$ conversion $\alpha_{\rm CO}$, this provides the first direct constraint on the total molecular gas density at $z>0$, tracing the reservoir over 12 Gyr. For comparison, we overlay estimates from CO galaxy surveys at $z = 0$ (FCRAO, xCOLD GASS)\cite{2003ApJ...582..659K,2021MNRAS.501..411F} and at higher redshifts (ASPECS, COLDz, HDF, PHIBSS2)\cite{2019ApJ...872....7R,2019ApJ...882..138D,2020AJ....159..190L,2020ApJ...902..110D,2023ApJ...945..111B}, the latter including only bright detected galaxies and thus shown as lower limits. At $z=1$--2.5, our total $\Omega_{\rm H_2}$ is about twice that resolved in galaxy surveys, revealing a substantial faint population previously missed and disfavoring shallow faint-end slopes. This comparison adopts the same $\alpha_{\rm CO}$ for all $z>0$ sources and intensity estimates, making their ratios independent of $\alpha_{\rm CO}$ assumptions. Early LIM results for CO from COPSS\cite{2016ApJ...830...34K} and mmIME\cite{2020ApJ...901..141K} support high $\Omega_{\rm H_2}$, though these detections are limited to small-scale shot power and require model-dependent conversion to mean CO. Two sets of upper limits are compatible with our result: a LIM constraint from COMAP\cite{2024A&A...691A.337C} and a null detection of CO absorbers in ALMACAL\cite{2019MNRAS.490.1220K}. Predictions from IllustrisTNG\cite{2019ApJ...882..137P}, EAGLE\cite{2015MNRAS.452.3815L}, and a semi-empirical model\cite{2015MNRAS.449..477P} are also shown, requiring no CO-to-H$_2$ conversion, with the latter two more consistent with our results.}
\label{fig:Omega_H2}
\end{figure}

\begin{figure} \centering
    \includegraphics[width=0.89\textwidth]{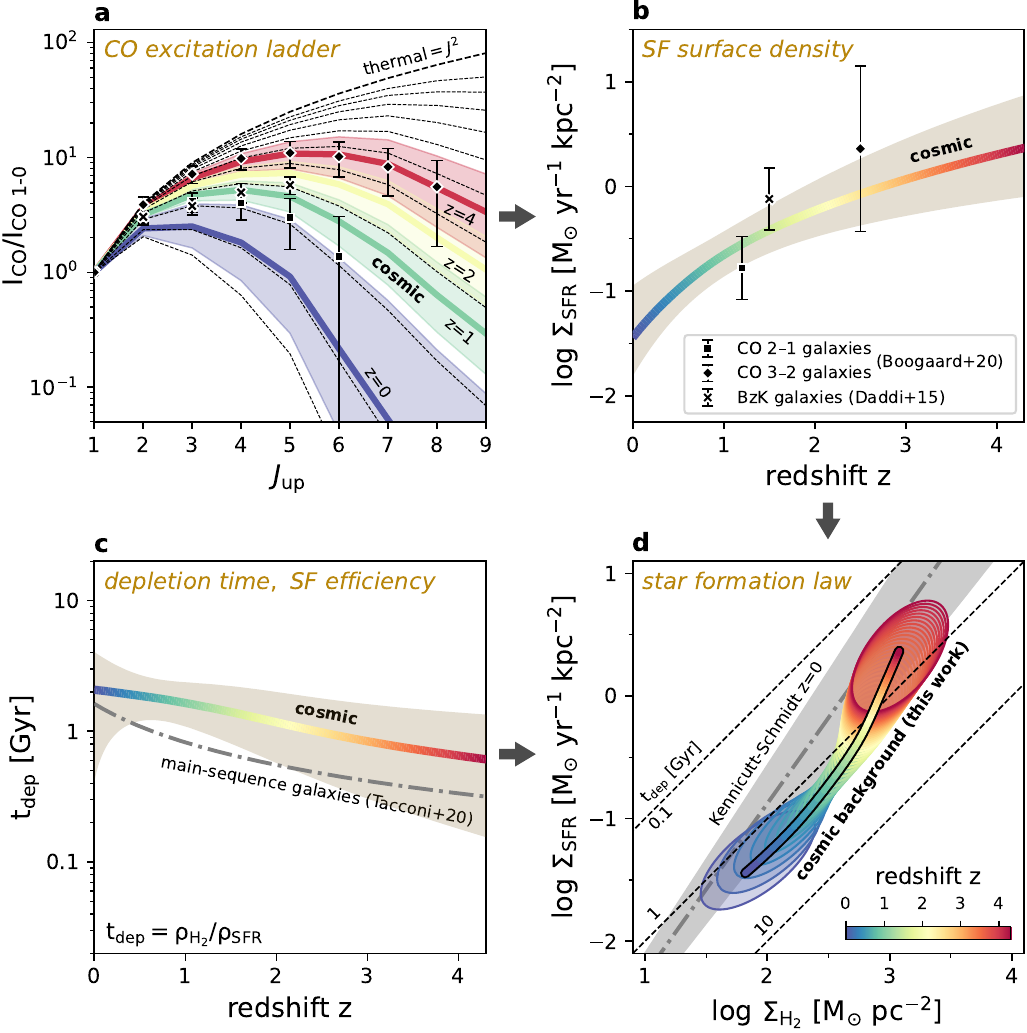}
    \vspace{-0.175cm}
    \caption{\textbf{Astrophysical summary of star-forming gas over cosmic history.} Thick solid lines and shaded bands or contours show our constraints from the cosmic CO background, colored consistently by redshift across all panels. Literature galaxy-based results are overplotted for comparison. All uncertainties are shown at the $1\sigma$ (68\%) level.
\textbf{a}, CO SLED, expressed as the velocity-integrated line intensity ratios versus $J$, revealing higher excitation at earlier epochs. 
\textbf{b}, Galaxy-scale star-formation surface density, $\Sigma_{\rm SFR}$, rising with redshift as constrained by the CO SLED in panel \textbf{a} (see \hyperref[sec_sled]{Methods}). 
\textbf{c}, Global molecular gas depletion time, $t_{\rm dep}=\rho_{\rm H_2}/\rho_{\rm SFR}$, derived from the CO-based $\rho_{\rm H_2}$ (Fig.~\ref{fig:Omega_H2}) and CIB-based $\rho_{\rm SFR}$\cite{2025ApJ...992...65C}, resembling the main-sequence galaxy relation\cite{2020ARA&A..58..157T}. A short $t_{\rm dep}$ relative to the Hubble time necessitates sustained inflow to maintain cosmic star formation. 
\textbf{d}, Star-formation law linking galaxy-scale surface densities, $\Sigma_{\rm SFR}\propto\Sigma_{\rm H_2}^{\sim1.5}$, 
derived from panels \textbf{b} and \textbf{c}. Color ellipses show $1\sigma$ uncertainties of the cosmic mean. The super-linear slope matches the local Kennicutt-Schmidt relation\cite{1998ApJ...498..541K}, revealing a universal mode of star formation that persists from galaxies to the aggregate background across 90\% of cosmic history.}
\label{fig:astro_summary}
\end{figure}

The CO background reveals the cosmic molecular history shown in Fig.~\ref{fig:Omega_H2}, tracing the fuel of cosmic star formation. This gas reservoir is dominated in mass by H$_2$, which is difficult to observe directly as it lacks a permanent dipole moment, but is traced by readily excited low-$J$ CO lines. Assuming a Milky-Way-like CO-to-H$_2$ conversion factor of $\alpha_{\rm CO}=3.6~{\rm M_\odot\,(K\,km\,s^{-1}\,pc^2)^{-1}}$\cite{2010ApJ...713..686D} for $J=1$--$0$, we infer the molecular density parameter $\Omega_{\rm H_2}\equiv\rho_{\rm H_2}/\rho_{\rm crit}$, where $\rho_{\rm H_2}=\alpha_{\rm CO}\,\rho_{\rm CO,10}$ is the comoving molecular gas density and $\rho_{\rm crit}$ the present-day critical density. Although the conversion is anchored at low-$J$ to extract mass rather than excitation, the constraint uses all transitions (Fig.~\ref{fig:astro_summary}a), with the SLED turnover well sampled across redshift (\hyperref[sec_line_sampling]{Methods}). We find that $\Omega_{\rm H_2}$ peaks at $z\sim1.5$ and declines toward both lower and higher redshifts, with evolution well described by:
\begin{eqnarray}
\Omega_{\rm H_2}(z) \approx 1.9\times10^{-4}\,\frac{(1+z)^{2.3}}{1+[(1+z)/2.8]^{6.5}}\ .
\label{eq:omega_H2_fit_function}
\end{eqnarray}
This delivers the first empirical $\Omega_{\rm H_2}$ from direct $\rho_{\rm CO}$ constraints, modulo $\alpha_{\rm CO}$, bypassing galaxy-survey incompleteness and avoiding the shot-noise-only extrapolation by measuring diffuse CO in the clustering regime.

Fig.~\ref{fig:Omega_H2} also overlays literature constraints on $\Omega_{\rm H_2}$ from integrating CO emission in detected galaxies, both locally\cite{2003ApJ...582..659K,2021MNRAS.501..411F} and at higher redshifts\cite{2019ApJ...872....7R,2019ApJ...882..138D,2020AJ....159..190L,2020ApJ...902..110D,2023ApJ...945..111B}. Although the absolute scale of $\alpha_{\rm CO}$ remains uncertain (see \hyperref[sec_alpha_CO]{Methods}), all $z>0$ studies adopt the same Milky-Way value used here, enabling direct comparison in their ratios. Beyond the local universe, only about 200 CO emitters of all $J$ have been detected across a total $<0.1~{\rm deg}^2$ area, sparsely distributed in redshift and leaving the faint end and sometimes even the knee of the evolving luminosity functions poorly constrained. As a result, these studies avoid extrapolation and report $\Omega_{\rm H_2}$ only from detected galaxies, which we plot as lower limits. During the peak star-formation epoch at $z=1$--2.5, our diffuse, wide-field CO intensity measurement yields a $\rho_{\rm CO,10}$ about twice that from individually detected galaxies above survey limits, giving an $\Omega_{\rm H_2}$ larger by the same factor under any shared $\alpha_{\rm CO}$ assumption. This factor-of-two gap most likely reflects incompleteness in galaxy surveys, revealing a substantial population of faint, previously missed galaxies. Combining individually detected bright populations with our total background estimates would yield constraints that disfavour top-heavy luminosity functions.

CO has been a key LIM target for probing 3D LSS via aggregate line emission. Alongside our results, Fig.~\ref{fig:Omega_H2} also summarizes existing LIM constraints, including the pioneering CO power-spectrum detections from COPSS\cite{2016ApJ...830...34K} and mmIME\cite{2020ApJ...901..141K} at 2--4$\sigma$, both limited to the shot-noise regime. These small-scale signals probe only the second moment of the luminosity function, making conversion to total CO emission (the first moment) and thus to $\Omega_{\rm H_2}$ highly model dependent. In addition, without redshift tomography, the single mmIME band suffers from line confusion, with a degeneracy among four different $J$'s and epochs. We also show two upper limits, both compatible with our detections: one from COMAP\cite{2024A&A...691A.337C}, a CO LIM experiment continuing to advance the constraints, and another from the non-detection of intervening CO absorbers\cite{2019MNRAS.490.1220K}, whose interpretation is less direct. Finally, we overlay three H$_2$ model predictions that do not require an $\alpha_{\rm CO}$ assumption: those obtained by post-processing the IllustrisTNG\cite{2019ApJ...882..137P} and EAGLE\cite{2015MNRAS.452.3815L} hydrodynamic simulations, together with a semi-empirical model\cite{2015MNRAS.449..477P} linking observed galaxy SFR to halo mass in $N$-body simulations. The latter two predict higher $\Omega_{\rm H_2}$ at $z=1$--3, in better agreement with our results.

We now exploit astrophysical information in the CO background for diagnostics of star-forming gas in the universe and summarize the results in Fig.~\ref{fig:astro_summary}. Beyond the overall amplitudes, we uncover a clear evolutionary trend in the relative strengths of the CO ladder, or the SLED, with rising excitation with redshift (Fig.~\ref{fig:astro_summary}a, see also Fig.~\ref{fig:spectrum_and_lines}b). In studies of individual galaxies, CO SLED varies strongly between normal galaxies, submillimeter-selected starbursts, and quasars\cite{2013ARA&A..51..105C}. By contrast, our measurement uniquely pinpoints the averaged excitation over cosmic time free from selection effects. The elevated SLED implies systematically higher galaxy star-formation surface densities $\Sigma_{\rm SFR}$ at high redshifts (Fig.~\ref{fig:astro_summary}b), consistent with denser molecular gas, stronger turbulence, and more intense radiation fields in the ISM\cite{2013ARA&A..51..105C,2020ARA&A..58..157T}.

The global molecular gas depletion time $t_{\rm dep}$ can be estimated by dividing our molecular
density $\rho_{\rm H_2}$ in Fig.~\ref{fig:Omega_H2} by the cosmic SFR density $\rho_{\rm SFR}$.
Fig.~\ref{fig:astro_summary}c shows the resulting posterior anchored on the recent $\rho_{\rm SFR}(z)$, derived from the CIB with a minor UV correction following equation~(42) of the companion study\cite{2025ApJ...992...65C}, for which we adopt a 10\% uncertainty.
Evaluated on a dense redshift grid, the evolution is well approximated by
\begin{eqnarray}
t_{\rm dep} \equiv \rho_{\rm H_2}/\rho_{\rm SFR} \simeq 3\,(1+z)^{-1}\ \rm Gyr\ ,
\label{eq:depletion_time}
\end{eqnarray}
on the long side but remains consistent with estimates for main-sequence galaxies\cite{2020ARA&A..58..157T}. This timescale of $\sim1$~Gyr is shorter than the Hubble time, so once gas cools into the molecular phase it is converted into stars, requiring ongoing replenishment through gas inflows that naturally occur alongside cosmological structure growth. Meanwhile, $t_{\rm dep}$ is far longer than the local free-fall time, indicating that some regulating processes, possibly feedback and turbulence, keep star-formation efficiency well below the dynamical limit\cite{2005ApJ...630..250K}. The redshift dependence of $t_{\rm dep}$ is mild compared with the strong evolution of both $\rho_{\rm H_2}$ and $\rho_{\rm SFR}$, implying that changes in the global fuel supply, rather than large variations in consumption efficiency, drive the cosmic star-formation history.

In the classic picture of star formation in galactic disks, surface densities regulate disk instability and the efficiency of gas collapse. We therefore investigate the star-formation law in Fig.~\ref{fig:astro_summary}d by comparing the characteristic galaxy-scale $\Sigma_{\rm SFR}$ from Fig.~\ref{fig:astro_summary}b inferred from CO excitation against the molecular gas surface density $\Sigma_{\rm H_2}=t_{\rm dep}\,\Sigma_{\rm SFR}$ obtained via depletion-time scaling. This assumes that $t_{\rm dep}$ derived from the volumetric ratios of $\rho_{\rm H_2}/\rho_{\rm SFR}$ in Fig.~\ref{fig:astro_summary}c can be applied to galaxy-scale surface quantities. Unlike galaxy-based scaling relations, the mean background yields only one point estimate in the $\Sigma_{\rm SFR}$-$\Sigma_{\rm H_2}$ plane at each redshift, with the fully propagated $1\sigma$ covariance also shown in Fig.~\ref{fig:astro_summary}d. As the universe evolves, these snapshots trace a cosmic star-formation law of the form $\Sigma_{\rm SFR}\propto\Sigma_{\rm H_2}^{N}$ with $N\approx1.5$, consistent with the local Kennicutt--Schmidt relation ($N\approx1.4$)\cite{1998ApJ...498..541K}. The super-linear slope indicates that higher molecular surface densities yield disproportionately more efficient consumption through enhanced gravitational cloud collapse\cite{2005ApJ...630..250K}. The persistence of this law, from local galaxies to global backgrounds and across 12~Gyr of cosmic history, suggests that the underlying physics may be fundamentally universal.

We now turn to [CII], the dominant coolant of diffuse ISM, excited by coupling stellar feedback in photodissociation regions, molecular cloud surfaces, and ionized medium. Each [CII] photon carries away thermal energy from collisions with gas particles, so the luminosity density $\rho_{\rm CII}$ measures the global cooling rate of star-forming gas after processes such as stellar winds, ionization, and photoelectric heating. From $\rho_{\rm CII}$ in Fig.~\ref{fig:spectrum_and_lines}b, we find a cooling rate evolution  
\begin{eqnarray}
\Lambda_{\rm CII}(z) \;\equiv\; \rho_{\rm CII}(z) 
\approx 5.9\times10^{38}\,\frac{(1+z)^{3.2}}{1+[(1+z)/2.9]^{6.6}}\ {\rm erg\ s^{-1}\ Mpc^{-3}}\ .
\label{eq:CII_cooling}
\end{eqnarray}
This provides a global benchmark for ISM energy balance, constraining the net feedback coupling of young stars to diffuse gas as an endpoint of the baryonic energy cascade, and complements the CO-based picture by explaining the cooling and funneling of cosmological inflows to replenish star-forming gas in cold clouds.

As [CII] responds directly to star formation and is the brightest line background, once calibrated it enables 3D LSS mapping through LIM, complementing CO. Using our [CII] detection at $z=2$, where $\rho_{\rm CII}$ is best constrained, we derive the mean [CII]--SFR relation:
\begin{eqnarray}
\rm {\it SFR} \approx 4.5^{+3.5}_{-1.6}\times 10^{-8}\ ({\it L}_{\rm CII}/{L}_{\odot})\ M_{\odot}\ {\rm yr}^{-1}\ ,
\label{eq:CII_SFR_relation}
\end{eqnarray}
where $L_{\rm CII}$ is the per-object [CII] luminosity. This is anchored to the precise CIB-based cosmic SFR density\cite{2025ApJ...992...65C}, same as that used for $t_{\rm dep}$, and assumes a Chabrier initial mass function\cite{2003PASP..115..763C}, providing the first global [CII] calibration at the peak star-formation epoch. Our scaling is consistent with local trends for most galaxy types\cite{2014A&A...568A..62D}, supporting a non-evolving [CII]--SFR relation. The ``[CII] deficit'' seen in ultraluminous infrared galaxies\cite{2013ApJ...774...68D} is not prominent in the [CII] background, reflecting the dominance of normal star-forming galaxies over rare starbursts in the cumulative light.

Together, these findings establish a coherent picture of galaxy formation: the molecular reservoir is larger than previously charted yet depleted on short timescales, requiring replenishment through cosmological matter assembly. CO excitation traces star-formation intensity under a universal, super-linear law set by gas supply, while [CII] captures the cooling imprints of stellar radiation and feedback coupling. Stellar growth in galaxies is thus sustained by a large, short-lived molecular supply, maintained by the continuous cycling of baryons via inflow, cooling-induced collapse, star formation, and feedback. This global, cosmological view anchors and contextualizes the diverse approaches to studying galaxy formation, from individual galaxies to cosmic backgrounds.

In the broader context of LIM, the sky background monopoles in Fig.~\ref{fig:T_b} and Extended Data Fig.~\ref{fig:I_nu} serve as anchor points for anisotropy signals, placing our CO and [CII] detections within a cosmological framework. At its core, the LIM approach expands LSS cosmology by using photons rather than galaxies as matter tracers. Unlike galaxy surveys, LIM does not impose surface brightness thresholds for source detection and requires much lower angular resolution, boosting survey efficiency, sky coverage, and completeness\cite{2017arXiv170909066K}. This enables access to LSS at higher redshifts with more linear modes, opening opportunities to probe reionization, dark energy, primordial non-Gaussianity, and beyond\cite{2019ApJ...872..126M,2022A&ARv..30....5B}. Yet before this work, results were limited to shot power\cite{2016ApJ...830...34K,2020ApJ...901..141K} and galaxy stacking\cite{2024arXiv240607861R}, both confined to small scales where line amplitudes do not generalize. The global strength of line backgrounds therefore remained unknown, leaving forecasts, sensitivity requirements, and survey strategies uncertain. Our detections of mean CO and [CII] provide the first empirical benchmark that directly closes this gap, establishing the line strengths needed to guide forecasts and instrument design across lines, frequencies, and redshifts.  

A growing set of experiments now targets CO and [CII]\cite{2014JLTP..176..767S,2018SPIE10708E..1OR,2020arXiv200914340V,2021JATIS...7d4004S,2022ApJ...933..182C,2022SPIE12190E..0QF,2023ApJS..264....7C}. Alongside our detections, Fig.~\ref{fig:T_b} overlays atmospheric transmission\cite{2001ITAP...49.1683P}, showing the frequency windows accessible from the ground, while a space mission could open the full parameter space\cite{2021ExA....51.1593S}. Our measurements provide empirical CO and [CII] line strengths and their evolution over a wide redshift range, closing the long-standing uncertainty that has limited LIM survey planning. Future experiments can therefore be designed and optimized using empirical targets rather than a broad span of model predictions. This establishes CO and [CII] as calibrated matter tracers, laying the empirical foundation for full 3D LIM measurements and robust cosmological inference, guiding LIM into its next phase of scientific development.

\clearpage

\clearpage
\newpage

\begin{methods}\label{sec_method}

\renewcommand{\figurename}{\bf Extended Data Fig.}
\setcounter{figure}{0}
\renewcommand{\tablename}{\bf Extended Data Table}
\setcounter{table}{0}

\subsection{Tomographic intensity mapping}\label{sec_tomo}

Our analysis employs clustering tomography, a method first applied to the UV background\cite{2019ApJ...877..150C} and later to the thermal Sunyaev-Zeldovich effect\cite{2020ApJ...902...56C} and the CIB\cite{2025ApJ...992...65C}, from which the intensity measurements used here are derived. The method cross-correlates diffuse intensity maps with spectroscopic reference samples, in this case 11 broadband maps from {\it Planck}, {\it Herschel}, and {\it IRAS}\cite{2005ApJS..157..302M,2012MNRAS.424.1614O,2016A&A...594A...8P,2017ApJS..233...26S} spanning 100\,GHz to 5\,THz, with $\sim$3 million galaxies and quasars from the Sloan Digital Sky Survey (SDSS)\cite{2002AJ....124.1810S,2015MNRAS.453.2779E,2016MNRAS.455.1553R,2017A&A...597A..79P,2020MNRAS.498.2354R,2021MNRAS.500.3254R}. On scales larger than individual dark matter halos, the fluctuations in intensity and galaxy density follow $\delta_I = b_I \delta_{\rm m}$ and $\delta_{\rm g} = b_{\rm g} \delta_{\rm m}$, where $b_I$ (denoted $b$ hereafter) and $b_{\rm g}$ are the respective bias factors. The tomographic cross-correlation amplitude measured in bins of redshift deprojects the extragalactic intensity $I$ without responding to foregrounds:
\begin{eqnarray}
\langle \delta_I \delta_{\rm g} \rangle \; = \; \frac{dI}{dz}\; b\, b_{\rm g}\, \langle \delta_{\rm m}^2 \rangle,
\label{eq:crosscorr}
\end{eqnarray}
where $dI/dz$ is intenisty $I$ emitted per unit redshift interval and $\langle \delta_{\rm m}^2 \rangle$ is the matter auto-correlation. Dividing by the theoretical $\langle \delta_{\rm m}^2 \rangle$ under a fiducial $\Lambda$CDM cosmology\cite{2020A&A...641A...6P} and correcting for the independently measured $b_{\rm g}$ yields empirical estimates of the background as the bias-weighted differential intensity $(dI/dz)b$. Because the 11 intensity maps have different beam sizes, the minimum physical scale cut ranges from 0.74 to 3.71~Mpc ($1.48$~Mpc for the most constraining bands), while the maximum is fixed at 14.84~Mpc. In this clustering regime, the reference galaxies’ own emission contributes only at the percent level or lower, ensuring robust estimates of $(dI/dz)b$. Adopting the cosmological radiative transfer solution in equation~(9) of the formalism\cite{2025ApJ...992...65C}, which is equivalent to a simple unit conversion, yields the bias-weighted CIB emissivity $\epsilon_\nu b$ in redshift-frequency space shown in Fig.~\ref{fig:spectrum_and_lines}a.

As described in the companion work\cite{2025ApJ...992...65C}, a small thermal Sunyaev-Zeldovich distortion from hot gas in the cosmic web, detected jointly at low frequencies\cite{2020ApJ...902...56C}, has already been removed from the CIB $(dI/dz)b$ data vector. Radio free-free and synchrotron emission are undetected even in our lowest-frequency band (100\,GHz), with their inclusion yielding only upper limits. We therefore do not include them in this work.

\subsection{Line sampling}\label{sec_line_sampling}

The intensity tomography $\epsilon_\nu b$ in Fig.~\ref{fig:spectrum_and_lines}a probes the total CIB, including both the dust continuum and spectral lines. Supplementary Fig.~\ref{fig:line_sampling} shows the coverage matrix for ionized carbon [CII] 158~$\mu$m and nine carbon monoxide CO transitions ($J=1$--0 through 9--8) across 16 redshift bins spanning $0<z<4.2$, based on instrument bandpasses and precise cross-correlation redshifts from spectroscopic galaxy references. Line coverages by {\it Planck} and {\it Herschel} are indicated separately. Of the 157 data points used for spectral fitting (below 8~THz rest-frame in Fig.~\ref{fig:spectrum_and_lines}a), 17 sample [CII]; 67 sample CO (31 include two transitions); the remaining 73 are in line-free regions, providing equally important continuum anchors for line detection.

CO lines from $J=4$--3 to 9--8 are fully sampled across $0<z<4.2$. Limited by the lowest {\it Planck} band (100\,GHz), lower-$J$ transitions are not directly covered at high redshifts. Nonetheless, the turnover behavior (Fig.~\ref{fig:astro_summary}a) of the CO excitation or the SLED is well sampled throughout, enabling robust simultaneous constraints on mass (low-$J$) and excitation (high-$J$). For [CII], coverage starts at $z\sim0.4$ and is complete at higher redshifts. This is faithfully reflected in the line luminosity posteriors (Fig.~\ref{fig:spectrum_and_lines}b), where the [CII] 68\% interval widens toward $z=0$.

\subsection{CO excitation}\label{sec_sled}
To link the nine CO transitions, we seek a family of SLEDs, $f_{\rm SLED}$, defined as the ratios of velocity-integrated line intensities $I$ or comoving luminosity densities $\rho$:
\begin{eqnarray}
f_{\rm SLED} \equiv I_{J(J{-}1)}/I_{10} = (\rho_{{\rm CO},J(J{-}1)}/\rho_{{\rm CO},10}) / J\ .
\label{eq:f_sled}
\end{eqnarray}
To parameterize $f_{\rm SLED}$ with a single variable controlling the level of excitation, we use results from numerical simulations of disk and merging galaxies with molecular-line radiative transfer\cite{2014MNRAS.442.1411N}, which identify the effective galaxy-scale star-formation surface density $\Sigma_{\rm SFR}$ as the primary regulator of CO excitation. Higher $\Sigma_{\rm SFR}$ corresponds to increased gas densities and temperatures, enhancing collisions between CO and H$_2$. Such conditions also entail stronger interstellar radiation fields, elevated cosmic-ray heating, and enhanced turbulence, all contributing to higher CO excitation. To capture this behavior, we adopt a modified version of the simulation-motivated functional form,
\begin{eqnarray}
\log f_{\rm SLED}(\Sigma_{\rm SFR}) = A\,\frac{\chi^B}{1+ \chi^B} + C\ ,\ \ {\rm with}\ \chi = \log\Sigma_{\rm SFR}-\chi_0\ ,
\label{eq:NK14}
\end{eqnarray}
where $I$ has units of Jy\,km\,s$^{-1}$, $\Sigma_{\rm SFR}$ is in $\mathrm{M_{\odot}\,yr^{-1}\,kpc^{-2}}$ with $\chi_0=-2$ set to the lowest $\log\Sigma_{\rm SFR}$ considered, and $A$, $B$, and $C$ are $J$-dependent coefficients. This form reduces to the original relation\cite{2014MNRAS.442.1411N} at $\chi^B\ll1$ and saturates at $\chi^B\gg1$, allowing a fully thermalized ceiling of $f_{\rm SLED}=J^2$ to be imposed with improved numerical stability. We set this thermalized bound at ${\rm log}\,\Sigma_{\rm SFR}=4$, although the exact anchor point is unimportant as long as it lies far above the $\Sigma_{\rm SFR}$ for typical starbursts.

The original simulated family of SLEDs was validated against available excitation measurements, especially for high $\Sigma_{\rm SFR}$ systems such as M82, submillimeter-selected dusty galaxies, and quasars, but was not well tested for more typical star-forming galaxies at lower $\Sigma_{\rm SFR}$ out to high $J$\cite{2014MNRAS.442.1411N}. For the cosmic CO background, however, the population is expected to be dominated by these more moderate, main-sequence-like systems\cite{2022A&A...667A.156B}, as commonly assumed in previous LIM studies\cite{2016ApJ...830...34K,2020ApJ...901..141K,2022ApJ...933..188B}, motivating an empirical recalibration. Multi-$J$ observations of normal star-forming galaxies beyond the local universe now exist for three samples at $z=1$--3: one “BzK” sample\cite{2015A&A...577A..46D} and two selected in CO $J=2$--1 and 3--2\cite{2020ApJ...902..109B}. Using their observed SLEDs and measured $\Sigma_{\rm SFR}$ values, we refit the coefficients in equation~(\ref{eq:NK14}) by requiring smooth evolution with both $J$ and $\Sigma_{\rm SFR}$, matching these observed SLEDs within uncertainties, and satisfying the thermalized boundary condition. The best-fit coefficients are listed in Extended Data Table~\ref{tab:SLED_coefficients}.

Based on this parametrization, Fig.~\ref{fig:astro_summary}a shows the allowed family of SLEDs across the full range of excitations (short dashed lines) compared to the observed ones for the galaxy samples at $z=1$--3\cite{2015A&A...577A..46D,2020ApJ...902..109B}, with their corresponding $\Sigma_{\rm SFR}$ shown in Fig.~\ref{fig:astro_summary}b. Relative to the original simulation output, our recalibration remains consistent at high $\Sigma_{\rm SFR}$ but yields a slower high-$J$ decline at low $\Sigma_{\rm SFR}$, in better agreement with the new measurements. This may reflect intrinsic $\Sigma_{\rm SFR}$ scatter or additional excitation mechanisms, both expected in the diverse population of galaxies contributing to the cosmic CO background. Unlike previous LIM studies with fixed SLEDs\cite{2016ApJ...830...34K,2020ApJ...901..141K,2022ApJ...933..188B}, our new parameterization, combined with the 11-band redshift tomography, enables a more data-driven inference of CO excitation alongside the attempt for first global CO background detection.

\subsection{Full spectral model}\label{sec_full_model}

Following the companion study\cite{2025ApJ...992...65C}, we adopt a spectral model that provides an analytic description of two observables: (1) the bias-weighted CIB emissivity $\epsilon_\nu b=\epsilon_\nu(\nu,z)\,b(z)$ (Fig.~\ref{fig:spectrum_and_lines}a); and (2) the CIB monopole $I_\nu(\nu_{\rm obs})$\cite{2019ApJ...877...40O} (cyan points in Supplementary Fig.~\ref{fig:line_monopole}), which constrains the black curve in Fig.~\ref{fig:T_b}. The latter is a derived quantity and provides an integral constraint on $\epsilon_\nu(\nu,z)$ after converting units to intensity. The CIB emissivity comprises thermal-dust continuum and line terms
\begin{equation}
\epsilon_\nu (\nu, z) = \epsilon_{\nu,\,\rm cont} (\nu, z) + \epsilon_{\nu,\,\rm line} (\nu, z)
\label{eq:total_emi}
\end{equation}
to be specified shortly. For the evolution of the light-weighted CIB bias factor, we adopt
\begin{equation}
b(z) = b_0 + b_1 z + b_2 z^2\ ,
\end{equation}
jointly constrained by the two observables $\epsilon_\nu b$ and $I_\nu$, one with bias, one without. 

The continuum term $\epsilon_{\nu,\,\rm cont}$ is parameterized as a generalized graybody spectrum\cite{2025ApJ...992...65C}, weighted by a range of dust properties to capture the diversity of galaxies contributing to the CIB:
\begin{eqnarray}
\epsilon_{\nu,\,\rm cont}(\nu,z)=\frac{4\pi\,\rho_{\rm d}(z)}{\Sigma_{\rm d}}\,[1-e^{-\langle\tau\rangle(\nu,z)}]\,\langle B_\nu\rangle(\nu,z)\ .
\label{eq:SED_1}
\end{eqnarray}
The key components are: $\rho_{\rm d}=\rho_{\rm d}(a,b,c,d)$, the cosmic dust mass density setting the overall normalization with 4 degrees of freedom (DOF) for redshift evolution; $\Sigma_{\rm d}$, the effective galaxy-scale dust surface density setting the transition wavelength between optically thin and thick regimes; $\langle\tau\rangle=\langle\tau\rangle(\beta_0,C_{\beta},\sigma_{\beta})$, the effective optical depth weighted by a distribution of dust opacity index $\beta$ with evolving mean and scatter; and $\langle B_\nu\rangle=\langle B_\nu\rangle(\mu_0,C_{\mu},s_T,\alpha_T)$, the Planck function weighted by a lognormal power-law dust-temperature distribution. Full parameterizations for each term are given in equation~(20), equations~(13) and (14), and equations~(17)--(19) of the companion paper\cite{2025ApJ...992...65C}, respectively.

We model the line term $\epsilon_{\nu,\,\rm line}$ from CO and [CII] simply by describing its unique frequency-redshift patterns: the emissivity excess in the $i$th data point, at redshift $z^i$ and rest-frame frequency $\nu^i$, sampled by a bandwidth $\Delta \nu^i = \Delta \nu_{\rm obs}^i (1 + z^i)$, is  
\begin{eqnarray}
\epsilon_{\nu,\, \rm{line}}^i &=& \epsilon_{\nu,\, \rm{CII}}^i + \epsilon_{\nu,\, \rm{CO}}^i \nonumber \\
&=& \frac{\rho_{\rm CII}(z^i)}{\Delta \nu^i} A_{\rm CII}^i + \sum_{\mathit{J}=1}^{9} \frac{\rho_{\rm CO,\, \mathit{J} \mathit{J}{-}1}(z^i)}{\Delta \nu^i} A_{{\rm CO},\, \mathit{J}}^i\,,
\label{eq:model_lines}
\end{eqnarray}
where $A_{{\rm CO},\, \mathit{J}}$ and $A_{\rm CII}$ are binary activation vectors, with elements set to 1 if the line falls within the rest-frame bandwidth and 0 otherwise (Fig.~\ref{fig:spectrum_and_lines}a and Supplementary Fig.~\ref{fig:line_sampling}). The nine CO lines are connected by CO SLEDs, $f_{\rm SLED}$, as parameterized in equations~(\ref{eq:NK14}) with coefficients in Extended Data Table~\ref{tab:SLED_coefficients}. We anchor CO (1--0) and [CII] to their local dust continua:
\begin{eqnarray}
\rho_{\rm{CO},10}(z) &=& R_{\rm{CO}}(z)\, \nu_{\rm{CO}}\,\epsilon_{\nu,\, \rm{cont}} (\nu_{\rm{CO}}, z)\,; \nonumber\\
\rho_{\rm{CII}}(z) &=& R_{\rm{CII}}(z)\, \nu_{\rm{CII}}\,\epsilon_{\nu,\, \rm{cont}} (\nu_{\rm{CII}}, z)\,,
\label{eq:CO_CII_norm_anchor}
\end{eqnarray}
where $\nu_{\rm{CO}}$ and $\nu_{\rm{CII}}$ are the rest frequencies for CO 1--0 and [CII], respectively, and $R_{\rm CO}(z)$ and $R_{\rm CII}(z)$ define the dimensionless equivalent widths quantifying the line-to-continuum ratios, to be constrained by the data. We allow $R_{\rm CO}$, $R_{\rm CII}$, and the CO SLED parameter $\Sigma_{\rm SFR}$ to evolve with redshift, each with one additional DOF, giving six free parameters in total for $\epsilon_{\nu,\,\rm{line}}$.

The full model has 21 free parameters. These include 12 for the CIB thermal dust continuum: $a$, $b$, $c$, $d$, $\mu_0$, $C_{\mu}$, $s_T$, $\alpha_T$, $\Sigma_{\rm d}$, $\beta_0$, $C_{\beta}$, and $\sigma_{\beta}$; 3 for the bias evolution: $b_0$, $b_1$, and $b_2$; and 6 for CO and [CII] lines: $R_{\rm{CO},\,0}$, $C_{\rm{CO}}$, $\log\,\Sigma_{\rm SFR,\,2}$, $C_{\Sigma_{\rm SFR}}$, $R_{\rm{CII,0}}^{100}$, and $C_{\rm{CII}}^{100}$. For convenience, in our fitting we use $R_{\rm{CII}}^{100} = 100 \times R_{\rm{CII}}$ to reflect the percent-level [CII]-to-continuum ratio expected, and adopt logarithmic forms for $a$, $\Sigma_{\rm d}$, and $\Sigma_{\rm SFR}$. Among these parameters, pairs of ($X_0$, $C_X$) describe the redshift evolution of each quantity $X \in \{\mu, \beta, R_{\rm CII}^{100}, R_{\rm CO}, \log\,\Sigma_{\rm SFR}\}$ related to the spectral shape:  
\begin{eqnarray}
X(z) = X_0 + C_X \log(1+z)\ .
\label{eq:par_redshift_evo}
\end{eqnarray}
Note that as long as no informative prior is imposed, the anchor redshift is mathematically arbitrary and does not affect the inferred $X(z)$ posteriors. This log-linear form is preferred over a power law for data-driven constraints, as it permits negative values (and sign flips) for noisy quantities such as $R_{\rm CII}^{100}$ and $R_{\rm CO}$, where line detections would otherwise be prior-driven if non-negativity were enforced. All $X$'s are anchored at $z=0$ via $X_0$, except for $\log\,\Sigma_{\rm SFR}$, which is anchored at $z=2$ with the same log-linear form 
$X(z) = X_2 + C_X[\log(1+z) - \log\,3]$, chosen only to test priors informed by observational results of CO excitation\cite{2015A&A...577A..46D,2020ApJ...902..109B} at cosmic noon. The full parameter set is summarized in Extended Data Table~\ref{tab:parameters}.

\subsection{Priors}\label{sec_prior}

Of the 21 model parameters, most could be constrained empirically, with only a few DOFs that are prior-driven, as summarized in Extended Data Table~\ref{tab:parameters}. For thermal dust and bias parameters, we adopt the same priors used before\cite{2025ApJ...992...65C}, largely uninformative except for a Gaussian prior on $\log \Sigma_{\rm d}$, which sets the optically thin-thick transition at frequencies above CO and [CII] and thus does not affect line detections. The evolving bias factor $b(z)$, assumed common to continuum and lines, is jointly constrained by $\epsilon_\nu b$ and $I_\nu$, with additional assumptions: a Gaussian prior $b_0=1\pm0.1$ for $z=0$, corresponding to dark matter halo masses of $M_{h}=10^{12.5}$--$10^{13}$~M$_\odot$\cite{2010ApJ...724..878T}, together with bounds $b(z=2)>2.4$ to exceed the bias of typical Lyman-break galaxies\cite{2005ApJ...619..697A} and $b(z=3)<4.81$ to remain below the extreme ${M_{h}}^{5/3}$ scaling relevant to the thermal Sunyaev-Zeldovich effect\cite{2020ApJ...902...56C}.  

For the line amplitudes $R_{\rm CO,0}$ and $R_{\rm CII,0}^{100}$ we impose only broad bounds so that the detections are data-driven. A weak Gaussian prior $C_{\rm CII}^{100}=0\pm2$ is applied to exclude extreme redshift evolution of [CII] relative to the continuum, while no such prior is placed on CO. For the CO SLED we adopt a Gaussian prior $\log\Sigma_{\rm SFR,2}=0.3\pm0.25$ at $z=2$, motivated by the galaxy samples used to calibrate the parameterization\cite{2015A&A...577A..46D,2020ApJ...902..109B} (see Fig.~\ref{fig:astro_summary}b), and require $\log\Sigma_{\rm SFR}>-2$ at all redshifts to avoid unphysical line ratios.

\subsection{Fitting to data}\label{sec_fitting}

We derive parameter posteriors using Bayesian inference, fitting the full continuum-plus-line model to the tomographic bias-weighted emissivity $\epsilon_\nu b$ and monopole $I_\nu$ data vector and covariance matrix. The exploration of parameter space is performed with Markov Chain Monte Carlo (MCMC) using the affine-invariant ensemble sampler implemented in the \texttt{emcee} package\cite{2013PASP..125..306F}. The resulting posteriors are listed in Extended Data Table~\ref{tab:parameters}, with the joint covariances between line and nuisance parameters shown in the segment of the full corner plot in Supplementary Fig.~\ref{fig:corner}.

The best-fit model projections in data space are shown as lines in Fig.~\ref{fig:spectrum_and_lines}a and Supplementary Fig.~\ref{fig:line_monopole}, which closely follow the data and provide a visual check of fit quality, especially for the continuum. Because CO and [CII] lines are much fainter, the overall goodness of fit is only modestly better than the continuum-only case in the companion study\cite{2025ApJ...992...65C}. Nonetheless, the reduced $\chi^2$ decreases from 1.49 to 1.40 despite six fewer DOFs, reflecting a measurable improvement, primarily from accounting for the low-frequency CO excess visible below $\sim$500~GHz in Fig.~\ref{fig:spectrum_and_lines}a.  

Both CO and [CII] are significantly detected (6.6$\sigma$ and 3.0$\sigma$) on top of the CIB continuum. The empirically recovered CO SLED trend in Fig.~\ref{fig:astro_summary}ab, showing increasing excitation with redshift, further supports the robustness of the CO detection. The 21 parameter posteriors fully describe the evolving CIB plus line spectrum and are propagated into derived quantities including the line luminosity densities (Fig.~\ref{fig:spectrum_and_lines}b), sky monopole brightness temperatures (Fig.~\ref{fig:T_b}), molecular gas history (Fig.~\ref{fig:Omega_H2}), and star-forming gas diagnostics (Fig.~\ref{fig:astro_summary}).

\subsection{Null test}\label{sec_null_test}

To verify that the CO and [CII] detections are not spurious, we perform a null test by repeating the MCMC fit 50 times after randomly shuffling the line-redshift associations in the activation vectors $A_{{\rm CO},\,J}$ and $A_{\rm CII}$ in equation~(\ref{eq:model_lines}). This destroys the physical frequency-redshift patterns while keeping all other inputs unchanged. The recovered line luminosity densities are consistent with zero, as shown in Supplementary Fig.~\ref{fig:null_test}, with dispersions smaller than the detections. This confirms that the signals arise only when the correct line placements are used.  

\subsection{Model-level preference}\label{sec_model_selection}
We now assess model selection together with parameter degeneracies. Because line luminosity densities are sub-percent relative to the CIB continuum, commonly used metrics like the Akaike and Bayesian Information Criteria (AIC/BIC) were not ideal for testing whether adding lines improved the fit, as they are ``amplitude-weighted'' and therefore strongly driven by the continuum. This means that continuum residuals at frequencies unrelated to the lines, e.g., near the CIB peak where the emissivity is highest, could saturate AIC and BIC and overshadow improvements from the faint emission lines. Nevertheless, adding lines does yield a modest reduction in $\chi^2$. Much stronger support for the model-level preference comes from several direct physical arguments discussed next.

Compared to the continuum-only fit\cite{2025ApJ...992...65C}, adding CO and [CII] leaves the posteriors for continuum parameters unchanged except for $\sigma_\beta$, the scatter in the dust opacity index $\beta$. Without lines, the low-frequency excess in Fig.~\ref{fig:spectrum_and_lines}a forces a large $\sigma_\beta\sim0.71$, broader than observed for diverse galaxy populations, and the $1\sigma$ lower tail extends to $\beta_{\rm low}\sim1$, in tension with galaxy observations\cite{2014PhR...541...45C}. With lines included, this excess is more correctly attributed to CO, $\sigma_\beta$ drops to $\sim0.24$, and the mean $\beta\sim2$, consistent with physical expectations. The continuum and lines are robustly separated with only a very weak degeneracy (correlation coefficient $-0.14$ between $\sigma_\beta$ and $R_{\rm CO,0}$ in Supplementary Fig.~\ref{fig:corner}), thanks to strong leverage from the multi-line CO pattern in redshift-frequency space. The continuum-plus-line model is therefore favored by the data and more physically realistic.

Moreover, the presence of CO and [CII] emission at known rest-frame frequencies is a direct physical expectation from quantum mechanics, so including them defines the more appropriate model space. In this sense, our inference of known lines is not subject to the ``look-elsewhere effect'' that would apply to blind searches over unknown frequencies. Within this physically preferred model, the data then constrain the line amplitudes, making the detections well grounded.

\subsection{Monopole}\label{sec_monopole}
We examine in detail the millimeter background monopoles from multiple LSS components, focusing on the CIB dust continuum and CO and [CII]. Supplementary Fig.~\ref{fig:line_monopole} presents these monopoles in intensity units, closely corresponding to Extended Data Fig.~\ref{fig:I_nu} and the brightness-temperature representation in Fig.~\ref{fig:T_b}, and adds direct comparisons with existing measurements and model predictions. As shown there, the 100--545\,GHz data points from {\it FIRAS} and {\it Planck} enter our spectral fit as integral constraints and are therefore consistent with the total-background posterior by construction. Supplementary Fig.~\ref{fig:line_monopole} further shows that our empirical CO amplitude, combining nine transitions, agrees with SIDES\cite{2022A&A...667A.156B}, a widely used model, but disfavors more extreme scenarios\cite{2024PhRvD.110b3513C} with suppressed low-$J$ amplitudes and steep high-$J$ decline. For [CII], our posterior is broadly consistent with recent predictions\cite{2022A&A...667A.156B,2022ApJ...929..140Y,2024PhRvD.110b3513C} but lies well below earlier claimed detections\cite{2018MNRAS.478.1911P,2019MNRAS.489L..53Y}, which were likely contaminated by correlated CIB continuum\cite{2019ApJ...872...82S}.

The CO background becomes increasingly important near and below 100\,GHz and may influence precision studies of CMB spectral distortion and secondary anisotropies\cite{2023PhRvD.107l3504M,2024PhRvD.110b3513C, 2025arXiv250616028M}. Toward the low-frequency end, the picture becomes more complex. Radio free-free and synchrotron emission rise as modeled in SIDES\cite{2022A&A...667A.156B} but remain negligible for our fit, which uses data above 100\,GHz. The thermal Sunyaev-Zeldovich decrement, extrapolated from empirical constraints at higher frequencies\cite{2020ApJ...902...56C}, becomes stronger than the other components below $\sim80$\,GHz in Supplementary Fig.~\ref{fig:line_monopole} (see also Fig.~\ref{fig:T_b} and Extended Data Fig.~\ref{fig:I_nu}). If the radio background catches up only slowly, this could produce a net negative LSS contribution, i.e., an ``absolute LSS decrement,'' in front of the CMB spanning several tens of GHz.

\subsection{Line to total infrared ratios}\label{sec_lines_to_total}

We evaluate the contributions of CO and [CII] to the total infrared luminosity density $\rho_{\rm TIR}$ in Supplementary Fig.~\ref{fig:line_over_TIR}a, finding that [CII] accounts for $\sim0.3$\% of $\rho_{\rm TIR}$ and CO for $\sim0.03$\%, in broad agreement with model expectations aside from some low-CO scenarios\cite{2024PhRvD.110b3513C}. This estimate is anchored to the $\rho_{\rm TIR}$ reported in equation~(39) of the companion study\cite{2025ApJ...992...65C}, for which we adopt a 10 percent uncertainty. Because the line contributions are small, the $\rho_{\rm TIR}$ from the continuum-only fit therein is nearly identical to that from the continuum-plus-line fit in this study. 

Supplementary Fig.~\ref{fig:line_over_TIR}a also shows that the line-to-total ratios rise mildly with redshift. The other panel demonstrates that this evolution becomes nearly flat when the line luminosity densities are normalized by $\rho_{\rm TIR}/f_{\rm obs}$, where $f_{\rm obs}$ is the cosmic dust-obscured fraction, which decreases toward higher redshift as UV emission from unobscured star formation becomes significant relative to the CIB\cite{2025ApJ...992...65C}. The resulting constant ratios of line to CIB+UV emission suggest that CO and [CII] trace the total SFR density more closely than the dust continuum. Producing CIB emission requires dust to be heated by young stars, but star formation alone does not ensure strong dust emission. In contrast, CO provides the fuel for star formation, and [CII] is excited as a direct consequence of it. These causal relations make the lines potentially more unbiased tracers of the total SFR in galaxy studies, especially in the early, dust-poor universe.

\subsection{Equivalent widths}\label{sec_equivalent_width}

An alternative way to characterize CO and [CII] line strengths is through their rest-frame equivalent widths relative to the local CIB continuum at the line frequency, rather than to the integrated energy budget. LIM noise is complex and often foreground-limited. If, however, residual CIB-continuum photon noise---an often overlooked component---dominates after standard high-pass filtering, the equivalent widths become especially informative, as they scale directly with the resulting signal-to-noise ratios.

To detect lines through their contrast with the continuum, our spectral framework naturally incorporates the dimensionless equivalent-width parameter $R$. We generalize $R$ in equation~(\ref{eq:CO_CII_norm_anchor}), originally defined for CO\,1--0 and [CII], to any line with rest frequency $\nu_0$, luminosity density $\rho_{\rm line}$, and underlying continuum emissivity $\epsilon_{\nu,{\rm cont}}(\nu_0)$:
\begin{eqnarray}
R \equiv \frac{\rho_{\rm line}}{\nu_0\,\epsilon_{\nu,{\rm cont}}(\nu_0)}\ .
\label{eq:R_generalized}
\end{eqnarray}
The conventional wavelength and frequency equivalent widths, $EW_\lambda$ and $EW_\nu$, then follow directly as
\begin{eqnarray}
EW_\lambda = R\,\lambda_0\,; \nonumber\\
EW_\nu = R\,\nu_0\,,
\label{eq:R_as_EW}
\end{eqnarray}
where $\lambda_0 = c/\nu_0$ and $c$ is the speed of light.

Using our empirically constrained 21-parameter posterior, we compute $R$, $EW_\lambda$, and $EW_\nu$ for the CO and [CII] backgrounds as functions of redshift in Supplementary Fig.~\ref{fig:EW}. This complements the line luminosity densities $\rho_{\rm line}$ in Fig.~\ref{fig:spectrum_and_lines}b. For CO, although $\rho_{\rm line}$ peaks at mid $J$, the equivalent widths increase monotonically toward low $J$ owing to the declining CIB continuum. Conversely, because [CII] lies near the thermal CIB peak, its high $\rho_{\rm line}$ still translates into substantially smaller equivalent widths than CO 1--0 by about 3.5 dex in $EW_\lambda$ and 1 dex in $EW_\nu$, with the midpoint of 2.25 dex (a factor of $\sim200$) in $R$. This underscores the added challenge for LIM experiments targeting [CII] and suggests that low-$J$ CO may be more accessible targets instead. As an additional note, this multi-line comparison also highlights the dimensionless $R$ as a more fundamental equivalent-width definition, free of the floating anchors inherent in wavelength or frequency units and thus well-suited for broader adoption in general astronomy applications.

\subsection{CO-to-H$_2$ conversion factor}\label{sec_alpha_CO}

The $\Omega_{\rm H_2}$ in Fig.~\ref{fig:Omega_H2} derived from CO 1--0 luminosity density (Fig.~\ref{fig:spectrum_and_lines}b) depends directly on $\alpha_{\rm CO}$. In the local universe, $\alpha_{\rm CO}$ is known to increase toward lower metallicity, especially below one-third solar, due to reduced dust and gas shielding of CO over H$_2$ against far-UV radiation\cite{2011ApJ...737...12L}. Even at fixed metallicity, the absolute calibration carries a 30~\% uncertainty\cite{2013ARA&A..51..207B}. Following common practice in high-redshift blank-field CO surveys (i.e., those that do not pre-select dusty starbursts) and in CO LIM studies, we adopt $\alpha_{\rm CO}=3.6~{\rm M_\odot\,(K\,km\,s^{-1}\,pc^2)^{-1}}$, consistent with the Milky Way value. Although an evolving $\alpha_{\rm CO}$ might be expected given the lower metallicities of higher-redshift galaxies, dynamical mass measurements of BzK galaxies at $z\sim1.5$ favor a non-evolving, Milky-Way-like value\cite{2010ApJ...713..686D}. This suggests either mild metallicity evolution in typical star-forming galaxies or ISM conditions that compensate for reduced relative shielding. Nevertheless, $\alpha_{\rm CO}$ remains a systematic uncertainty in estimating $\Omega_{\rm H_2}$.

Despite these caveats, relative amplitudes of $\Omega_{\rm H_2}$ in Fig.~\ref{fig:Omega_H2} are robust, as most data points adopt the same $\alpha_{\rm CO}$; the ratios are thus unaffected by this assumption and correspond to ratios of the underlying CO\,1--0 luminosity densities. Two exceptions are xCOLD GASS\cite{2021MNRAS.501..411F} at $z=0$, which employs a metallicity-dependent $\alpha_{\rm CO}$, and the ALMACAL CO absorber-based limits\cite{2019MNRAS.490.1220K}, which rely on a column-density version of the conversion factor. With a consistent $\alpha_{\rm CO}$ applied to all other measurements, our main finding that intensity mapping reveals roughly twice the molecular gas mass already resolved in galaxies holds directly. Any future revision of $\alpha_{\rm CO}$ would simply rescale all inferred $\Omega_{\rm H_2}(z)$ in Fig.~\ref{fig:Omega_H2} but leave their ratios unchanged.

A varying $\alpha_{\rm CO}$ would, however, affect the inferred depletion time $t_{\rm dep}$ and the star-formation law in Fig.~\ref{fig:astro_summary}c and d. A higher $\alpha_{\rm CO}$ at early times would imply larger H$_2$ reservoirs for the same CO intensity, yielding longer $t_{\rm dep}$ and a shallower cosmic star-formation law deviating from the Kennicutt-Schmidt relation. Conversely, if one assumes that cosmic star formation follows the Kennicutt-Schmidt law with a fixed slope near $N\approx 1.4$, consistent with our best-fit $N\approx 1.5$, then the effective $\alpha_{\rm CO}$ would evolve only weakly with redshift, in line with our fiducial assumption.

\end{methods}

\begin{addendum}
\item[Data availability]
The clustering-based tomographic CIB data vector and covariance matrix from the companion study\cite{2025ApJ...992...65C} used for the joint continuum-plus-line constraints in this work are publicly available on Zenodo (\url{https://zenodo.org/records/16486649})\cite{chiang_2025_16486649}. The posterior CO and [CII] line luminosity densities, along with the derived sky monopole intensities and brightness temperatures, are released on Zenodo (\url{https://zenodo.org/records/15495143})\cite{chiang_2025_15495143}. The MCMC posterior chains for all model parameters are released together with this manuscript.

\item[Code availability]
The custom code used for the CIB-plus-line fitting will be made publicly available upon journal publication. The MCMC analyses were performed using the \texttt{emcee}\cite{2013PASP..125..306F} package. Standard scientific Python libraries including \texttt{numpy}, \texttt{scipy}, \texttt{astropy}, and \texttt{matplotlib} were used throughout the analysis.

\end{addendum}

\begin{addendum}
 
 \item[Acknowledgements] 
 We acknowledge discussions with C. Hirata on line-signal extraction, G. K. Keating on model expectations, Y.-N. Lee and N. Yoshida on the interpretation of the star-formation law, and Q. D. Wang, B. M\'enard, U.-L. Pen, L.-H. Lin, J. Vieira, and T. Wong for helpful comments on the manuscript. Y.-K.C. is supported by the National Science and Technology Council of Taiwan through grants NSTC 111-2112-M-001-090-MY3 and NSTC 114-2112-M-001-063-MY3, and by Academia Sinica through the Career Development Award AS-CDA-113-M01.

 \item[Author contributions] Y.-K.C. conceived the study, performed all analyses, and wrote the manuscript.

 \item[Competing interests] The author declares no competing interests.

\end{addendum}

\clearpage

\renewcommand{\figurename}{\bf Extended Data Fig.}
\setcounter{figure}{0}
\renewcommand{\tablename}{\bf Extended Data Table}
\setcounter{table}{0}

\begin{table}
    \centering
    \caption{CO SLED coefficients}
    \begin{threeparttable}
    \makebox[\textwidth][c]{%
        \begin{tabular}{cccc}
            \hline\hline
            $\rm transition$ & $A$ & $B$ & $C$ \\
            \hline
            $J=2-1$ & $0.3$ & $1.95$ & $0.31$ \\
            $J=3-2$ & $0.86$ & $1.52$ & $0.15$ \\
            $J=4-3$ & $1.52$ & $1.42$ & $-$$0.2$ \\
            $J=5-4$ & $2.26$ & $1.46$ & $-$$0.71$ \\
            $J=6-5$ & $3.36$ & $1.57$ & $-$$1.62$ \\
            $J=7-6$ & $4.66$ & $1.47$ & $-$$2.66$ \\
            $J=8-7$ & $6.56$ & $1.19$ & $-$$4.05$ \\
            $J=9-8$ & $8.7$ & $0.98$ & $-$$5.51$ \\
            \hline\hline
        \end{tabular}
    }
    \caption*{Recalibrated coefficients for equation~(\ref{eq:NK14}), constrained using CO observations of star-forming galaxies at $z=1$--3\cite{2015A&A...577A..46D,2020ApJ...902..109B} and the thermalized maximum excitation limit. Together, these define the parameterization for the allowed family of SLEDs used in our CO analysis as shown in Fig.~\ref{fig:astro_summary}ab.}
    \end{threeparttable}
    \label{tab:SLED_coefficients}
\end{table}

\begin{table*}
    \centering
    \caption{Summary of model parameters}
    \scalebox{0.985}{
    \begin{threeparttable}
    \begin{tabular}{llrlr}
		\hline\hline
parameter & meaning & \multicolumn{2}{l}{$\ \ \ \ \ $range and prior} & posterior \\ 
\hline
log~$a$ & dust density normalization & [4, 6] & flat & $5.01^{+0.11}_{-0.10}$ \\ 
$b$ & dust density evolution & [0, 4] & flat & $1.58^{+0.31}_{-0.30}$ \\
 $c$ & dust density evolution & [0, 5] & flat & $2.84^{+0.14}_{-0.14}$ \\ 
 $d$ & dust density evolution & [4, 9] & flat & $6.31^{+0.38}_{-0.35}$\\ 
 $\mu_0$ & logarithmic dust temperature at $z=0$ & [2, 3] & flat & $2.57^{+0.08}_{-0.08}$\\ 
 $C_{\mu}$ & logarithmic dust temperature evolution & [$-1$, 2] & flat & $0.63^{+0.10}_{-0.10}$\\ 
 $s_T$ & logarithmic dust temperature spread & [0, 0.2] & flat & $0.05^{+0.06}_{-0.04}$\\ 
 $\alpha_T$ & temperature distribution high-end index & [4, 6] & flat & $5.18^{+0.21}_{-0.18}$\\ 
 log~$\Sigma_{\rm d}$ & effective galaxy dust surface density & [7, 8.5] & $\mathcal{N}(7.7,\,0.1)$ & $7.61^{+0.10}_{-0.10}$\\ 
 $\beta_0$ & mean opacity spectral index at $z=0$ & [1.3, 3] & flat & $2.21^{+0.14}_{-0.14}$\\ 
 $C_{\beta}$ & mean opacity spectral index evolution & [$-1.5$, 1.5] & flat & $0.14^{+0.24}_{-0.23}$\\ 
 $\sigma_{\beta}$ & opacity spectral index spread & [0.1, 1.5] & flat & $0.24^{+0.13}_{-0.10}$\\ 
 \hline
 $b_0$ & effective CIB bias at $z=0$ & [0, 1.5] & $\mathcal{N}(1,\,0.1)$ &  $0.95^{+0.10}_{-0.10}$\\ 
 $b_1$ & effective CIB bias evolution & [0.5, 1.5] & flat & $0.65^{+0.08}_{-0.09}$\\ 
 $b_2$ & effective CIB bias evolution & [0, 1] & flat & $0.04^{+0.05}_{-0.03}$\\ 
\hline
 $R_{\rm{CO},\,0}$ & CO line-to-continuum ratio at $z=0$ & [$-5$, 5] & flat & $1.05^{+0.91}_{-0.83}$\\ 
 $C_{\rm{CO}}$ & CO line-to-continuum ratio evolution & [$-20$, 20] & flat & $1.51^{+3.69}_{-3.09}$\\ 
 log~$\rm \Sigma_{SFR,\, 2}$ & CO SLED SFR surface density at $z=2$ & [$-2$, 3] & $\mathcal{N}(0.3,\,0.25)$ & $-0.23^{+0.28}_{-0.25}$\\ 
 $C_{\rm{\Sigma SFR}}$ & SFR surface density evolution & [$-2$, 6] & flat & $2.52^{+1.01}_{-1.19}$\\ 
 $R_{\rm{CII,0}}^{100}$ & [CII] line-to-continuum ratio at $z=0$ & [$-10$, 10] & flat & $0.55^{+0.85}_{-0.85}$\\ 
 $C_{\rm{CII}}^{100}$ & [CII] line-to-continuum ratio evolution & [$-10$, 10] & $\mathcal{N}(0,\,2)$ & $1.18^{+1.72}_{-1.73}$\\ 
\hline\hline
\end{tabular}
    \caption*{Parameter descriptions, priors, and posteriors, grouped into three blocks (top to bottom): CIB thermal dust continuum, clustering bias factor, and CO and [CII] line emission. Additional bounds on the combined bias evolution are described in \hyperref[sec_prior]{Methods}.}
    \end{threeparttable}}
    \label{tab:parameters}
\end{table*}

\begin{figure*} \centering
    \includegraphics[width=1\textwidth]{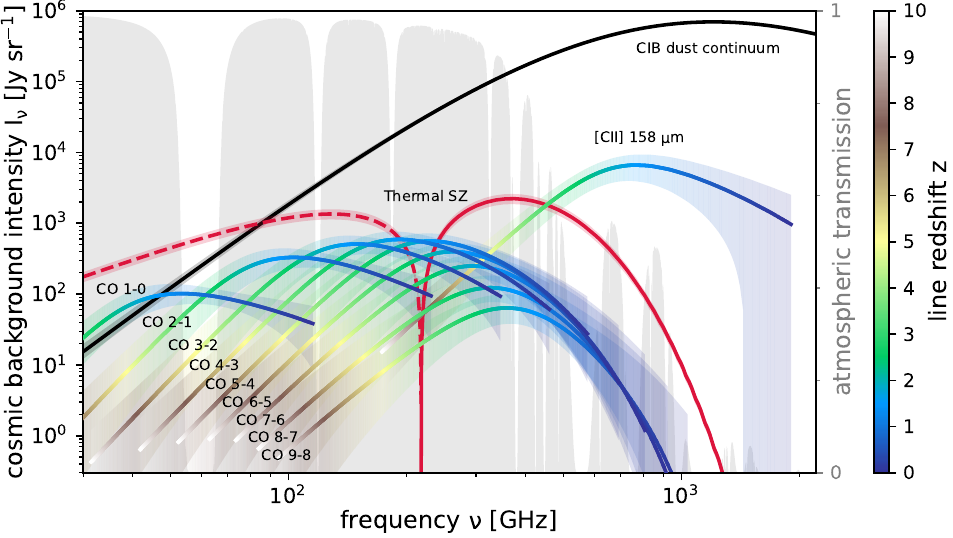}
    \caption{\textbf{Landscape of millimeter line monopole intensities.} Same as Fig.~\ref{fig:T_b}, but shown in specific intensity $I_\nu$ (Jy sr$^{-1}$) rather than brightness temperature. Curves show posterior median CO and [CII] line monopoles, color-coded by redshift. The jointly fitted CIB dust-continuum monopole is shown in black, and the sum with the line contribution matches the CIB-only fit in the companion work\cite{2025ApJ...992...65C}, based on the same data vector. The thermal Sunyaev–Zeldovich spectral distortion is shown in red, with the dashed segment indicating the decrement\cite{2020ApJ...902...56C}. All shaded bands correspond to 68\% uncertainties. Filled gray regions show atmospheric transmission windows at a dry site\cite{2001ITAP...49.1683P}. This figure is analogous to Supplementary Fig.~\ref{fig:line_monopole} but includes only updated empirical constraints from cross-correlation-based tomographic intensity mapping.}
\label{fig:I_nu}
\end{figure*}

\clearpage

\providecommand{\JournalTitle}[1]{#1}
\providecommand{\bibinfo}[2]{#2}
\providecommand{\bibfield}[2]{#2}
\providecommand{\BibitemOpen}{}
\providecommand{\BibitemShut}[1]{}

\clearpage

\section*{Supplementary Information}%
\setlength{\parskip}{12pt}%

\renewcommand{\figurename}{\bf Supplementary Fig.}
\setcounter{figure}{0}
\renewcommand{\tablename}{\bf Supplementary Table}
\setcounter{table}{0}

\vfill
\begin{figure}[h]
    \centering
    \includegraphics[width=0.575\textwidth]{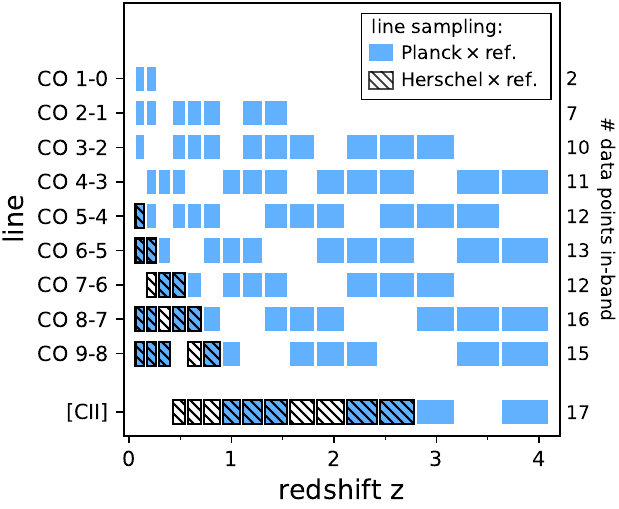}
    \caption{\textbf{Sampling of CO and [CII] lines over redshift.} The matrix shows whether a CO or [CII] line contributes, at a given redshift, to a data point in the bias-weighted CIB emissivity (Fig.~\ref{fig:spectrum_and_lines}a), measured from tomographic cross-correlations with spectroscopic reference objects\cite{2025ApJ...992...65C}. Blue-filled and black-hatched cells mark coverage by {\it Planck} and {\it Herschel} bands, respectively. The number of data points sampling each line is listed at right. Together, these provide sufficient coverage to track CO and [CII] emission, along with the CIB continuum, over 12~Gyr.}
\label{fig:line_sampling}
\end{figure}
\vfill

\begin{figure}
    \centering
    \includegraphics[width=0.65\textwidth]{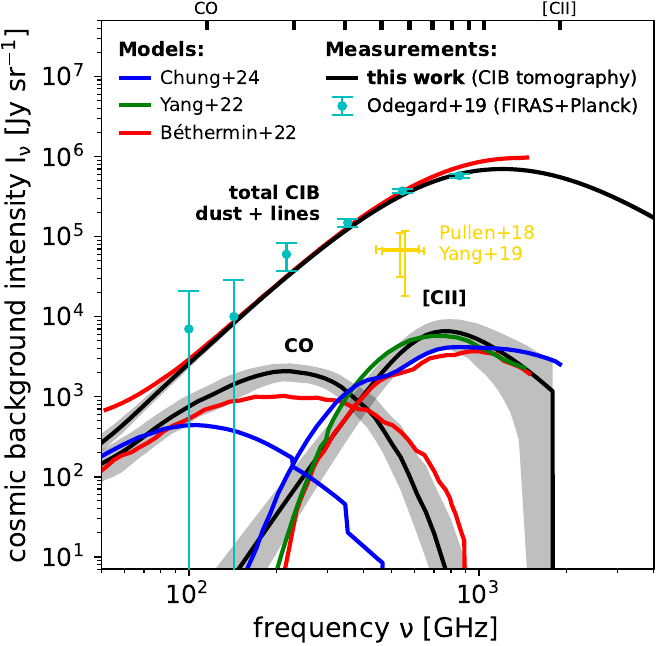}
    \caption{\textbf{Contribution of CO and [CII] to the CIB monopole.} Sky-averaged intensities of the total CIB (dust continuum plus lines), CO (nine transitions combined), and [CII] are shown as functions of observed frequency, complementing the brightness-temperature view in Fig.~\ref{fig:T_b}. Black lines and shaded bands show our posterior medians with 68\% credible intervals. Cyan markers denote total CIB monopoles (1$\sigma$) from {\it FIRAS} and {\it Planck}\cite{2019ApJ...877...40O}, which we use as external integral constraints. Earlier claims of exceptionally high [CII] at $z\sim2.6$ (yellow, $2\sigma$)\cite{2018MNRAS.478.1911P,2019MNRAS.489L..53Y} lie closer to the total CIB than to plausible [CII] amplitudes. Our empirical line measurements yield CO and [CII] constraints that broadly agree with two recent models\cite{2022A&A...667A.156B,2022ApJ...929..140Y} while disfavouring another CO prediction\cite{2024PhRvD.110b3513C}. Although [CII] has a higher absolute intensity, CO becomes increasingly important relative to the CIB continuum toward lower frequencies, underscoring its relevance for precision CIB and CMB analyses.}
\label{fig:line_monopole}
\end{figure}

\begin{figure}
    \centering
    \includegraphics[width=1\textwidth]{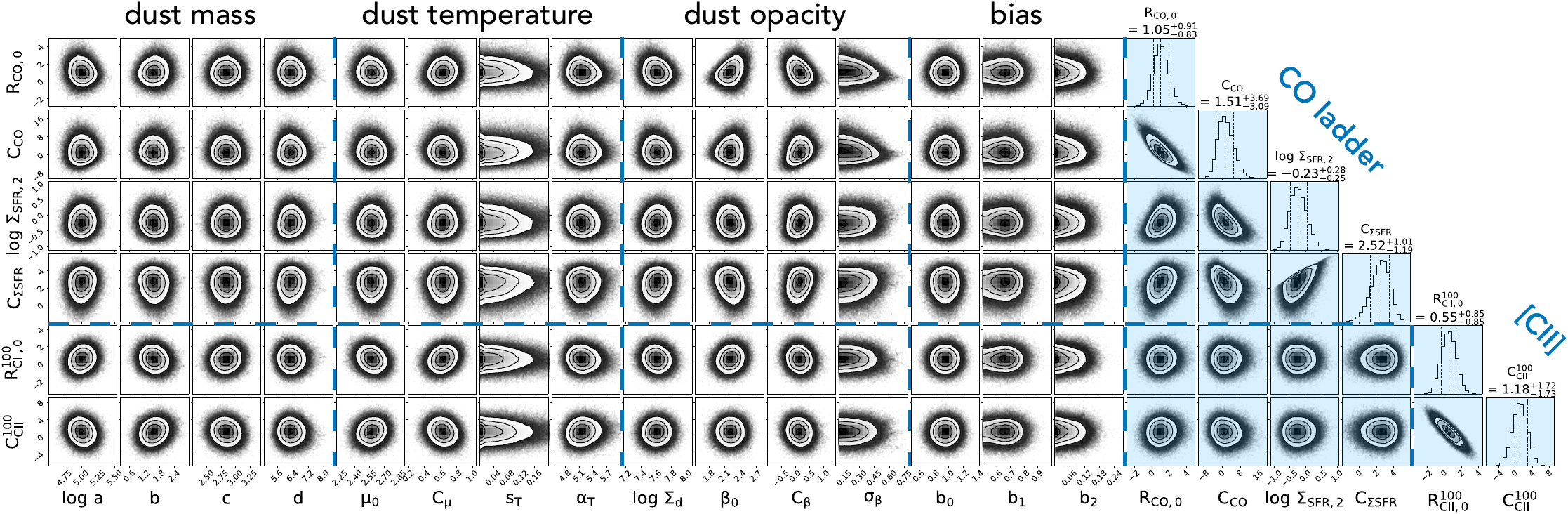}
    \caption{\textbf{Posteriors and parameter degeneracy.} Marginalized 1D posteriors of the six CO and [CII] parameters are shown together with their 2D joint posteriors with the other fifteen continuum parameters, extracted from the full 21-parameter MCMC corner plot. The continuum parameters are grouped into those describing the evolution of the CIB in dust mass, temperature, opacity index, and effective bias. As expected from their distinct spectral signatures, the line parameters exhibit no strong correlations or degeneracies with the continuum, indicating that the line signals are cleanly and unambiguously isolated.}
\label{fig:corner}
\end{figure}

\begin{figure} 
	\centering
        \includegraphics[width=0.5\textwidth]{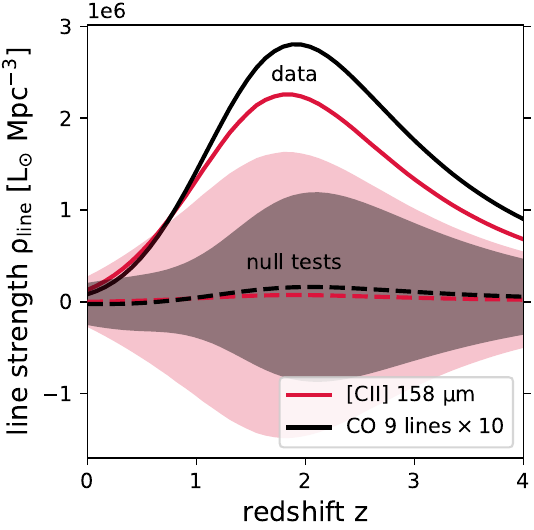} 
        \caption{\textbf{Randomized null tests of CO and [CII] histories.} To test the robustness of our line detections, 50 realizations of the full 21-parameter fit to data are performed with shuffled line mapping in the vectors $A_{{\rm CO},\,J}$ and $A_{\rm CII}$. This breaks the frequency-redshift correspondence and therefore should yield non-detections. Dashed lines and shaded bands show the mean and 68\% scatter of the resulting null-test posteriors for line luminosity densities versus redshift, compared with the solid curves from the detections (Fig.~\ref{fig:spectrum_and_lines}b). CO amplitudes, summed over nine lines, are scaled by a factor of ten for visibility. The null-test results are consistent with zero, confirming that the signals appear only when the correct line pattern is used.}
\label{fig:null_test}
\end{figure}

\begin{figure} 
	\centering
        \includegraphics[width=0.95\textwidth]{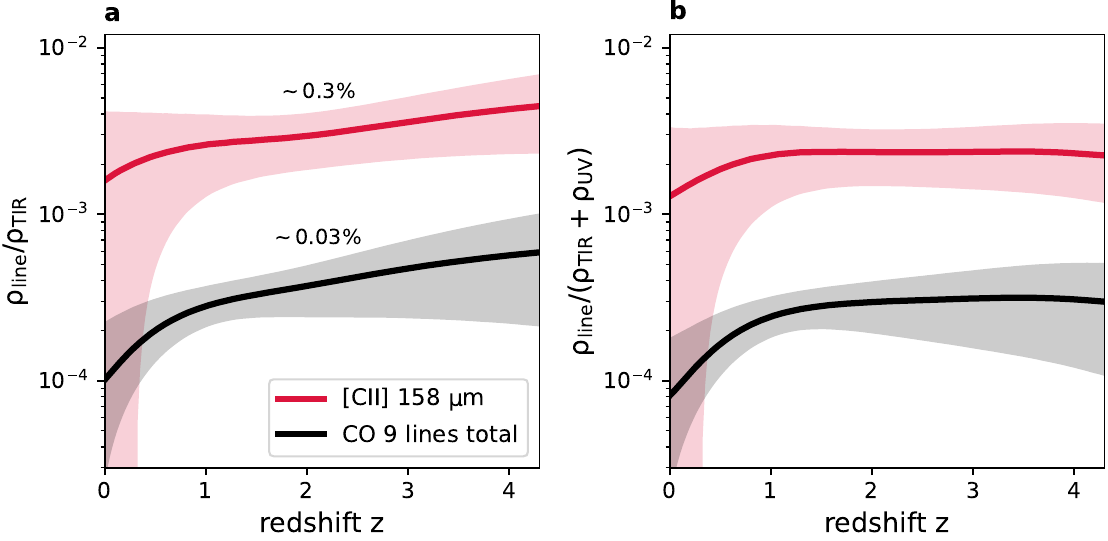}
        \caption{\textbf{Evolution of line to total CIB ratios.} \textbf{a}, Fractional contributions of cosmic [CII] and CO luminosity densities (Fig.~\ref{fig:spectrum_and_lines}b) to the CIB TIR luminosity density\cite{2025ApJ...992...65C}, with lines showing posterior medians and shaded regions indicating 68\% intervals. We find levels of $\sim0.3\%$ for [CII] and $\sim0.03\%$ for the nine CO lines combined, both showing weak redshift evolution. \textbf{b}, The same line luminosity densities normalized by the combined TIR-plus-UV luminosity density, where the UV component corrects the modest incompleteness of TIR in tracing the total cosmic star-formation history\cite{2025ApJ...992...65C}. These corrected ratios appear closer to being constant with redshift, indicating that [CII] and CO may be more unbiased tracers of the total SFR than the CIB, especially at high redshift.}
\label{fig:line_over_TIR}
\end{figure}

\begin{figure} 
	\centering
        \includegraphics[width=1\textwidth]{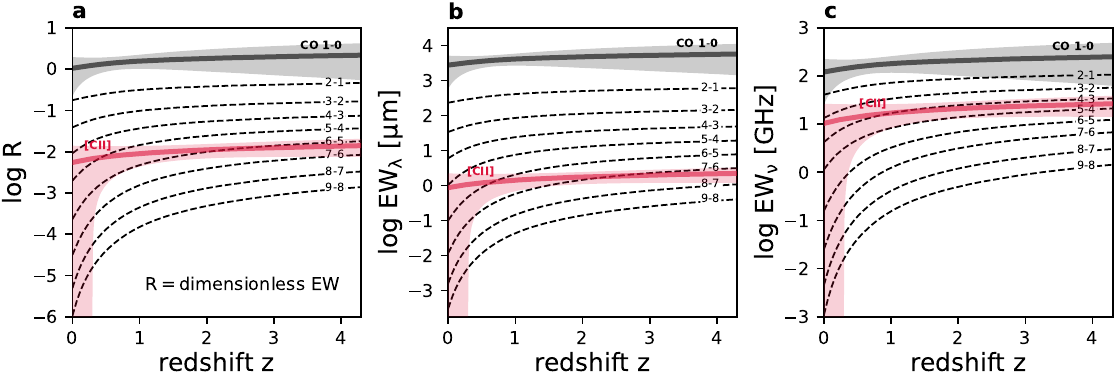}
        \caption{\textbf{Line equivalent width evolution for cosmic CO and [CII].} \textbf{a}, Dimensionless line-to-continuum ratios $R$ for CO transitions (black) and [CII] (red) as functions of redshift, with lines showing posterior medians and shaded regions indicating 68\% intervals. \textbf{b}, \textbf{c}, Corresponding wavelength and frequency equivalent widths, $EW_{\lambda}$ and $EW_{\nu}$, derived from $R$. All curves are computed from our full intensity-tomography posterior and illustrate the line-continuum contrast at the respective line frequencies. Low-$J$ CO lines exhibit substantially higher equivalent widths than [CII], a consideration that complements absolute line strengths when forecasting LIM detectability.}
\label{fig:EW}
\end{figure}

\end{document}